\documentclass[lettersize,journal]{IEEEtran}
\usepackage{amsmath,amsfonts}
\usepackage{algorithmic}
\usepackage{array}
\usepackage[caption=false,font=normalsize,labelfont=sf,textfont=sf]{subfig}
\usepackage{textcomp}
\usepackage{stfloats}
\usepackage{url}
\usepackage{verbatim}
\usepackage{graphicx}

\def\BibTeX{{\rm B\kern-.05em{\sc i\kern-.025em b}\kern-.08em
    T\kern-.1667em\lower.7ex\hbox{E}\kern-.125emX}}
\usepackage{balance}

\newcommand{\mrm}[1]{\mathrm{#1}}
\newcommand{\at}{{\,\vert\,}}

\usepackage{bm}
\usepackage{nicefrac}
\usepackage{tabularx}

\begin{document}
\title{Embedded Model Predictive Control for EMS-type Maglev Vehicles}
\author{%
Arnim Kargl, Mario Hermle, Zhiqiang Zhang, Yanmin Li, Dainan Zhao, Yong Cui, Peter Eberhard
\thanks{This work did not receive external funding.}
\thanks{A. Kargl, M. Hermle, and P. Eberhard: Institute of Engineering Computational Mechanics, University of Stuttgart, Pfaffenwaldring~9, 70569~Stuttgart, Germany. (e-mail: {arnim.kargl@itm.uni-stuttgart.de}; {mario.hermle@itm.uni-stuttgart.de}; {peter.eberhard@itm.uni-stuttgart.de})}
\thanks{Z. Zhang, Y. Li, and D. Zhao: State Key Laboratory of High-speed Maglev Transportation Technology, CRRC Qingdao Sifang Co., Ltd., No.~88 Jinhongdong Road, Chengyang District, 266111~Qingdao, People's Republic of China. (e-mail: {zhangzhiqiang@cqsf.com}; {liyanmin@cqsf.com}; {zhaodainan@cqsf.com})}
\thanks{Y. Cui: Chinese-German Research and Development Centre for Railway and Transportation Technology Stuttgart (CDFEB e.V.), Torstraße~20, 70173~Stuttgart, Germany. (e-mail: {yong.cui@ievvwi.uni-stuttgart.de})}
\thanks{This work has been submitted to the IEEE for possible publication.
Copyright may be transferred without notice, after which this version may no longer be accessible.
}
}

\markboth{}{Kargl, Hermle, \MakeLowercase{\textit{(et al.)}: Embedded Model Predictive Control for EMS-type Maglev Vehicles}}

\maketitle

\begin{abstract}
Current developments of high-speed magnetic levitation technology using the principle of the electromagnet suspension (EMS) focus on reaching vehicle speeds of more than 600\,km/h.
With increasing vehicle speeds, however, updated control algorithms need to be investigated to reliably stabilize the system and meet the demands in terms of ride comfort.
This article examines the modern and popular approach of model predictive control and its application to the magnetic levitation control system.
Investigated key aspects are the parameterization of the model predictive controller and its implementation on embedded, resource constrained hardware.
The results reveal that model predictive control is capable to robustly stabilize the highly nonlinear and constrained system even at very high speed.
Furthermore, processor-in-the-loop studies are carried out to validate the designed control algorithms on a microcontroller.
\end{abstract}


\section{Introduction}
\label{sec:intro}
\IEEEPARstart{T}{he} continuing growth of mobility in society combined with the aspect of globalization requires a fast, efficient, and sustainable mode of transportation.
Hereby, high-speed magnetic levitation (Maglev) vehicles have the potential to bridge the existing gap between traditional wheel-on-rail systems and air travel~\cite{HanKim16,WenkEtAl18}.
The contact-less design principle of a Maglev vehicle reduces wear as well as noise emissions and allows to achieve much higher speeds than classical railroad systems.
Current research at the State Key Laboratory of High-speed Maglev Transportation Technology in China together with the Chinese-German Research and Development Center for Railway and Transportation Technology in Stuttgart and the University of Stuttgart~\cite{DingEtAl23} focuses on the development of a high-speed Maglev train with a speed of more than $600\,\mrm{km/h}$, providing a future alternative to cover medium-range distances, replacing short-distance flights.

The Maglev system in development is based on the electromagnetic suspension (EMS) principle, which means that the vehicle is lifted and guided with attractive magnet forces, generated by electromagnets distributed along the train~\cite{HanKim16}.
\begin{figure}[b]
	\centering
	\includegraphics{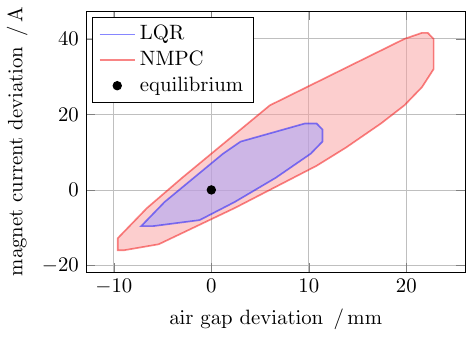}
	\caption{Regions of attraction of the Maglev system using an LQR and nonlinear MPC.}
	\label{fig:regions_of_attraction}
\end{figure}
The propulsion of the vehicle is realized as long-stator linear synchronous motor.
Since the EMS represents an unstable system, automatic control for both the levitation system and guidance system is necessary to stabilize a given equilibrium point.
During the early stages of the Transrapid's design in Germany, classical linear control techniques were studied and successfully implemented for this purpose~\cite{GottzeinMeisingerMiller80}.
However, the increasing demands on the control system, aiming for higher vehicle speeds and enhanced ride comfort, cannot be met by simple linear controllers.
This becomes clear when introducing the highly nonlinear system dynamics as well as corresponding state and input constraints in Section~\ref{sec:maglev_system}.
To illustrate the limitations of linear control, Figure~\ref{fig:regions_of_attraction} shows the regions of attraction (ROA) of the closed-loop Maglev system formulated in Section~\ref{sec:maglev_system}, comparing a linear quadratic regulator (LQR) with the nonlinear model predictive controller (MPC) from Section~\ref{sec:mpc_design}.
The ROA is defined as the set of all system states from which the controller is able to steer the system back to the desired equilibrium point.
The regions visualized are computed using numerical simulations of the closed-loop system for a dense grid of initial conditions in the shown state space.
It can be seen that the LQR controller is only able to stabilize a small region around the equilibrium point.
This is expected, since the LQR is based on a linearized model, which is only valid in a small neighborhood around the equilibrium point.
In contrast, nonlinear MPC is able to take into account the full nonlinear dynamics of the system, resulting in a larger region of attraction, especially regions where nonlinear effects dominate the system behavior.
Thus, MPC allows to operate the Maglev system safely under more challenging conditions, e.g., at higher vehicle speeds or with larger disturbances acting on the system.

For these reasons, current research focuses on the application of model predictive control~\cite{RawlingsMayneDiehl17} to the Maglev system.
MPC is an advanced, optimization-based control technique for general nonlinear multi-input-multi-output systems, which, e.g., can directly incorporate constraints on the system's state or input into the control law.
First studies on nonlinear MPC in the Maglev context are reported in~\cite{SchmidEberhardDignath19}.
In~\cite{SchmidEtAl21}, accurate and at the same time fast to evaluate magnet models are presented, which are efficiently implementable in an MPC context.
More recent developments contain offset-free control of the vehicle's levitation system~\cite{SchmidEberhard21}, controlling the lifting and lowering of the vehicle~\cite{HermleKarglEberhard25}, or studies concerning special scenarios, such as magnet failure~\cite{KarglEtAl25}.
Throughout the development, TR.Mechatron, a powerful simulation toolbox, is established~\cite{DignathEtAl25}, which amongst others, includes highly detailed multibody vehicle models~\cite{SchneiderEtAl21} for analysis purposes.

This contribution focuses on the development and application of embedded MPC algorithms for the magnetic levitation system of the Maglev vehicle.
The focus is set on the levitation system, since it is the most important aspect of EMS-type Maglev control.
Section~\ref{sec:maglev_system} introduces the magnetic levitation system and its mathematical modeling for control purposes.
Section~\ref{sec:mpc_design} presents the design of the model predictive controller, the respective optimal control problem, and the overall control architecture.
The implementation and solution of the optimal control problem on embedded, resource constrained hardware is discussed in Section~\ref{sec:mpc_implementation}.
Finally, Section~\ref{sec:results} illustrates simulation and processor-in-the-loop results before concluding in Section~\ref{sec:conclusion}.

\section{Magnetic Levitation System}
\label{sec:maglev_system}
\begin{figure*}[t]
	\centering
	\includegraphics{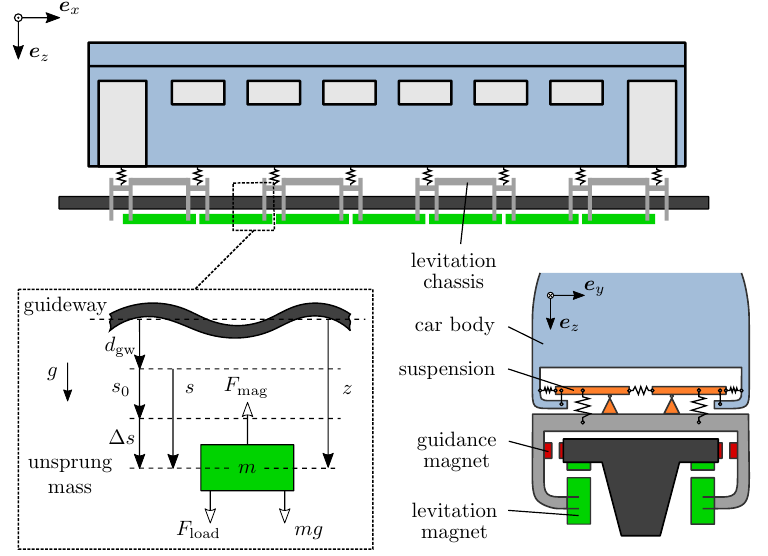}
	\caption{Cross-sections of an EMS-type magnetic levitation vehicle and illustration of the electromagnetically levitated single-point mass.}
	\label{fig:maglev_system}
\end{figure*}

The key components of an EMS-type Maglev vehicle are the magnetic levitation system as well as the guidance system.
Both systems use electromagnets distributed along the train to lift the vehicle and guide it along the guideway, respectively, compare Figure~\ref{fig:maglev_system}.
Sensors placed on each magnet provide relevant measurements for control, such as the air gap between magnet and guideway or the electric current through the respective electromagnet.
The control loop is closed by two control units per levitation magnet, which provide the input voltage to the electromagnets based on the measured signals to stabilize the system at a desired set point.
Figure~\ref{fig:maglev_system} highlights the most important components of the magnetic levitation and guidance system.
In the following, the focus is set to the magnetic levitation system, since it is the more challenging system to control, especially at high vehicle speeds.
The reason for this is the construction of the guideway in form of a regularly pillared track.
The elasticity of the individual girders, their unevenness, and finally offsets from one girder to the next~\cite{ZhengEtAl18} are the main causes for disturbances acting on the levitation control system.

For control purposes, one usually makes use of the so-called \textit{principle of the magnetic wheel}, as proposed in~\cite{GottzeinMeisingerMiller80}.
In essence, the mechanical structure of the Transrapid vehicle is set up such that individual electromagnets are almost decoupled from each other.
This is part of the overall vehicle's safety concept and allows for independent control of each electromagnet, i.e., of each half of a standard levitation magnet.
The stability of the complete system is then inferred from the stability of each controlled subsystem.
Hence, the control task simplifies to the magnetically levitated mass point, as sketched in Figure~\ref{fig:maglev_system}.
The resemblance to a single wheel suspension justifies the name \textit{magnetic wheel}~\cite{GottzeinMeisingerMiller80}.

The relevant system dynamics for one half of a standard levitation magnet can be described as follows.
The absolute position of the magnetically levitated mass point is denoted by 
\begin{equation}
	\label{eq:half_magnet_absolute_position}
	z(t) = s(t) + d_\mrm{gw}(t)
\end{equation} 
with respect to a fixed frame of reference, where $s(t)$ is the air gap and $d_\mrm{gw}(t)$ the deflection of the guideway.
The differential equation for the mechanical domain follows directly from Newton's second law, namely,
\begin{equation}
	\label{eq:half_magnet_mechanics}
	\ddot{z}(t) = g + \frac{F_{\mrm{load}}(t) - F_{\mrm{mag}}(t)}{m}
	\;.
\end{equation}
Hereby, $g$ denotes the gravitational constant, $m$ is the considered mass, $F_{\mrm{load}}$ corresponds to the vehicle's load, i.e., due to the distributed weight of the car body, and $F_{\mrm{mag}}$ is the magnetic force, which is provided in more detail later.
Dependencies on the time~$t$ are explicitly mentioned here to distinguish constants from time-dependent variables.
However, note that a dependency on the time does not directly imply any information about the timescale.
For example the load force~$F_{\mrm{load}}$ changes on a slow timescale, e.g., due to varying passenger numbers or aerodynamic effects, compared to the magnetic force~$F_{\mrm{mag}}$.
Changes in the magnetic force are limited to the electromagnet's dynamics discussed next.
Following the derivation in~\cite{SchmidEtAl21}, said dynamics can be described using the equivalent magnetic circuit method, which results in a high dimensional differential algebraic equation (DAE).
It provides a detailed model, including the effects of magnetic saturation, eddy currents, and fringing fluxes often neglected in Maglev control.
However, the provided DAE description is not necessarily usable for model predictive control purposes, especially, concerning real-time control.
Therefore, the major contribution in~\cite{SchmidEtAl21} is a numerically derived, simplified model, which maps the high dimensional DAE onto a nonlinear ordinary differential equation (ODE) with just one electric current as state variable.
During this process, which may be interpreted as a model order reduction, the loss of precision is negligible.
The resulting model can be formulated by the ODE
\begin{equation}
	\label{eq:half_magnet_magnet_ode}
	\dot{I}(t) = \alpha\big(s(t), \dot{s}(t), I(t)\big) \, I(t) + \beta\big(s(t), I(t)\big) \, U(t)
\end{equation}
in the state~$I$, which denotes the electric current.
The voltage~$U$ serves as input to the system, determined by the controller, and the terms~$\alpha$ and $\beta$ collect the nonlinearities.
Note that on the real hardware, the maximum allowed voltage which can be applied to the electromagnet is limited, such that $U(t) \in [U_{\mrm{min}}, U_{\mrm{max}}]$.
The magnetic force follows from the same numerical model order reduction in form of a lookup table, i.e.,
\begin{equation}
	\label{eq:half_magnet_magnet_force}
	F_{\mrm{mag}}(t) = F_{\mrm{mag}}(s(t), I(t))
	\;.
\end{equation}

\subsection{Analysis Model}
Equations~\eqref{eq:half_magnet_absolute_position}--\eqref{eq:half_magnet_magnet_force} combined form the model of the magnetically levitated mass, which is used for \textit{analysis} purposes.
The real-world system contains sensors placed on the electromagnet to measure the air gap signal~$s(t)$.
In simulation, the air gap~$s(t)$ is computed as the difference between the absolute position~$z(t)$ and the guideway deflection~$d_{\mrm{gw}}(t)$ following Equation~\eqref{eq:half_magnet_absolute_position}.
Setting up the state vector with $\bm{x}_{\mrm{an}} := [z \; \dot{z}\; I]^{\top}$ yields the state-space description 
\begin{equation}
\begin{aligned}
	\label{eq:analysis_model_dynamics}
	\dot{\bm{x}}_\mrm{an}(t) &= \bm{f}_\mrm{an}(\bm{x}_\mrm{an}, u, \bm{d})  \\ 
	&= 
	\begin{bmatrix}
		\dot{z} \\[0.5em]
		g + \frac{F_{\mrm{load}} - F_{\mrm{mag}}(z, d_{\mrm{gw}}, I)}{m} \\[0.5em]
		\alpha(z, \dot{z}, d_{\mrm{gw}}, \dot{d}_{\mrm{gw}}, I) I + \beta(z, d_\mrm{gw} I) u
	\end{bmatrix}
\end{aligned}
\end{equation}
of the analysis model in absolute coordinates.
The control input is the applied voltage~$u=U$ to the electromagnet.
For control purposes the output collects the air gap signal, the magnet's absolute acceleration, and the electric current, i.e.,
\begin{equation}
	\label{eq:analysis_model_output}
	\bm{y}_{\mrm{an}}(t) = \bm{h}_\mrm{an}(\bm{x}_\mrm{an}, \bm{d}) =
	\begin{bmatrix}
		s \\ \ddot{z} \\ I
	\end{bmatrix}
	=
	\begin{bmatrix}
		z - d_{\mrm{gw}} \\[0.5em]
		g + \frac{F_{\mrm{load}} - F_{\mrm{mag}}(z, d_{\mrm{gw}}, I)}{m} \\[0.5em]
		I
	\end{bmatrix}
	\;,
\end{equation}
which corresponds to the measured outputs of the real system.
The load force consists of a nominal force~$F_{\mrm{load},0}$ and a disturbance force~$d_\mrm{load}$, i.e., $F_{\mrm{load}} = F_{\mrm{load},0} + d_\mrm{load}$.
Together with the guideway irregularities, the disturbance input to the analysis model is given by $\bm{d} := [d_{\mrm{gw}} \; \dot{d}_{\mrm{gw}} \; d_\mrm{load}]^{\top}$.
These disturbances cannot be influenced by the controller, but have to be compensated to ensure stable and offset-free levitation of the vehicle.

In the aforementioned simulation toolbox TR.Mechatron \cite{DignathEtAl25}, the model's components are implemented in Matlab and finally connected in a Simulink model, which provides a modular approach and enables quick exchange of individual components.
Furthermore, TR.Mechatron provides much more detailed vehicle models for analysis purposes, such as a longitudinal cross-section model of three complete vehicle sections, travelling on an elastic guideway~\cite{SchneiderEtAl21}.

\subsection{Synthesis Model}
For \textit{controller synthesis} tasks, the system dynamics is slightly reformulated.
Firstly, the control objective is to follow a desired reference for the air gap signal~$s$, rather than setting the magnet's absolute position~$z$.
From a control perspective, the guideway's deflection is an unknown disturbance and assumed to be zero, i.e., $d_{\mrm{gw}} \equiv 0$, which corresponds to the nominal case. 
The disturbance rejection is therefore achieved through feedback compensation.
Hence, the state $z$ can be directly replaced with the air gap $s$.
A similar argument could be stated for the unknown changes to the load force~$F_{\mrm{load}}(t)$.
However, as stated earlier, these changes happen on a slow timescale compared to the magnet dynamics or guideway irregularities.
Following the discussion in~\cite{SchmidEberhard21}, the load force is not simply replaced by the nominal load force~$F_{\mrm{load},0}$, but the disturbance~$d_\mrm{load}$ is estimated such that
\begin{equation}
	\widehat{F}_{\mrm{load}}(t) = F_{\mrm{l},0} + \widehat{d}_{F_\mrm{l}}(t) 
	:= F_{\mrm{l},0} +k_\mrm{s} \int_{0}^{t} s(\tau) - s_0 \;\mrm{d}\tau
	\;,
\end{equation}
which guarantees offset-free tracking of the nominal air gap~$s_0$.
The constant~$k_\mrm{s}$ influences the speed of the estimation and is tuned during control design.

Finally, the system dynamics is shifted, such that the origin becomes the equilibrium point to be stabilized, i.e., it holds $\bm{f}(\bm{0}, 0) = \bm{0}$.
The original variables are related to the shifted variables by $s = s_0 + \Delta s$, $I = I_0 + \Delta I$, and $U = U_0 + \Delta U$.
Hence, an equilibrium point~$(s_0, I_0, U_0)$ is computed and the synthesis model is formulated with state
$\bm{x} := [\Delta s \; \dot{s} \; \Delta I]^\top$ and input~$u := \Delta U$ to
\begin{equation}
	\label{eq:synthesis_model_ode}
	\dot{\bm{x}}(t) =
	\bm{f}(\bm{x},u) =
	\begin{bmatrix}
		\dot{s} \\[0.5em]
		g + \frac{\widehat{F}_{\mrm{load}} - F_{\mrm{mag}}(\Delta s, \Delta I)}{m} \\[0.5em]
		\alpha(\Delta s, \dot{s}, \Delta I) \Delta I + \beta(\Delta s, \Delta I) \Delta u
	\end{bmatrix}
	\;.
\end{equation}
The synthesis model's outputs remain the same as the analysis model's, but are expressed in the shifted variables through
\begin{equation}
	\label{eq:synthesis_model_output}
	\bm{y}(t) = \bm{h}(\bm{x}) =
	\begin{bmatrix}
		s \\ \ddot{z} \\ I
	\end{bmatrix}
	=
	\begin{bmatrix}
		s_0+\Delta s \\[0.5em]
		g + \frac{\widehat{F}_{\mrm{load}} - F_{\mrm{mag}}(\Delta s, \Delta I)}{m} \\[0.5em]
		I_0 + \Delta I
	\end{bmatrix}
	\;.
\end{equation}
At this point it is worth noting that the synthesis model's outputs do not have to reflect the measured outputs of the real system.
In fact, in case of set-point stabilization in state space, no output equation is necessary at all.
However, the MPC control design in Section~\ref{sec:mpc_design} makes use of the output equation to define the control objective in terms of output references, which is also known as set-point stabilization in the output space.
This has the advantage that not only the air gap, but also the magnets' acceleration can be intuitively tuned through the cost function.

\section{Model Predictive Control Design}
\label{sec:mpc_design}
With the model description in Section~\ref{sec:maglev_system}, the formulation of the model predictive control problem follows in a standard fashion.
Since the \textit{design} of the MPC problem is largely decoupled from the actual \textit{implementation}, the two aspects are covered separately.

\begin{figure}[t]
	\centering
	\includegraphics{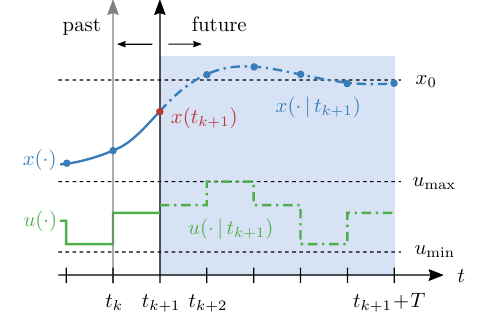}
	\caption{Receding horizon scheme visualized for a dynamic system with scalar state and input.}
	\label{fig:mpc_receding_horizon}
\end{figure}

The general idea of model predictive control can be described with the visualization in Figure~\ref{fig:mpc_receding_horizon}.
Given a state measurement $x(t_k)$ at time $t_k$, an open-loop optimal control problem~(OCP) is solved to determine the optimal sequence of inputs, which result in stabilization of the origin.
Part of solving the optimal control problem involves predicting the system's behavior into the future over a finite time horizon by utilizing its internal dynamic model, hence the name model \textit{predictive} control.
The resulting optimization minimizes a cost functional that typically penalizes tracking errors and control effort, while enforcing the internal model dynamics as well as state and input constraints. 
To compute the optimal input trajectory~$u^*(\cdot \,\vert\, t_k)$, the OCP is solved at each sampling time step~$t_k=k\delta$ with $k\in\mathbb{N}_0$ and sampling period~$\delta$ over a finite prediction horizon~$T \in \mathbb{R}$.
Subsequently, only the first part of the computed input trajectory~$u_\mrm{MPC}(t_k) = u^*(t_k,\vert\, t_k)$ is applied to the plant, which closes the control loop.
Note that the applied input is held constant over the sampling period~$\delta$, i.e., a zero-order hold is implemented.
At the next sampling time~$t_{k+1}$, the current state~$x(t_{k+1})$ is measured again, and the whole procedure is repeated, resulting in the so-called receding horizon principle.

The formulation of the OCP for the levitation control of one half magnet directly follows from the synthesis model in Equations~\eqref{eq:synthesis_model_ode} and~\eqref{eq:synthesis_model_output} and is given by
\begin{subequations}
\label{eq:ocp_levitation}
\begin{align}
	\label{eq:ocp_levitation_a}
	& \hspace{-3.6cm}\underset{u(\cdot \at t_k)}{\mrm{minimize}} \int_{t_k}^{t_k + T} \bm{y}^\top(t \at t_k) \bm{Q} \bm{y}(t \at t_k) + R u^2(t \at t_k) \;\mrm{d}t \\
	\label{eq:ocp_levitation_b}
	\mrm{subject~to} \quad \bm{x}(t_k \at t_k) &= \hat{\bm{x}}(t_k)\;, \\
	\label{eq:ocp_levitation_c}
	\dot{\bm{x}}(t \at t_k) &= \bm{f}\big(\bm{x}(t \at t_k), u(t \at t_k)\big) \;, \\
	\label{eq:ocp_levitation_d} 
	u(t \at t_k) &\in \mathcal{U}
\end{align}
\end{subequations}
with $t \in [t_k, \; t_k + T)$. Equation~\eqref{eq:ocp_levitation_a} defines the cost function to be minimized, where $\bm{Q} = \operatorname{diag} (Q_s,\, Q_{\ddot{z}},\, Q_I) \succeq 0$ and $R > 0$ are weighting matrices for the output and input, respectively.
The initial condition for prediction is set in Equation~\eqref{eq:ocp_levitation_b} with the current state~$\hat{\bm{x}}(t_k)$ at time~$t_k$ taken from direct measurement or a state estimator.
To predict the future system behavior, the internal model dynamics~\eqref{eq:synthesis_model_ode} are enforced as a constraint in Equation~\eqref{eq:ocp_levitation_c}.
Finally, input constraints are considered in Equation~\eqref{eq:ocp_levitation_d}, where $\mathcal{U} \in [u_\mrm{min}, \; u_\mrm{max}]^\top = [U_\mrm{min}-U_0, \; U_\mrm{max}-U_0]^\top$ defines the set of admissible inputs.
Note that for numerical accuracy, the state~$\bm{x}$, input~$u$, and the output~$\bm{y}$ are scaled linearly before being used in the OCP formulation.
However, the scaling is not explicitly shown in the OCP~\eqref{eq:ocp_levitation} for the sake of clarity.
Furthermore, no ingredients to formally guarantee closed-loop stability, such as terminal costs or terminal constraints, are included in the OCP.
The synthesis model~\eqref{eq:synthesis_model_ode} contains lookup tables, which makes the derivation of stability proofs based on terminal ingredients difficult or impossible.
Furthermore, a terminal region introduces an additional state constraint, which may lead to infeasibility of the OCP in practice.
Thus, a sensible length for the prediction horizon $T$ is determined in simulative investigations to ensure closed-loop stability at least in an empirical sense.

The MPC scheme is based on state feedback, which requires accurate state estimation. 
To this end, the measured outputs~\eqref{eq:analysis_model_output} are available to provide an estimate~$\hat{\bm{x}}(t_k)$ of the current state at each sampling time~$t_k$.
Evidently, the air gap~$s$ and the electric current~$I$ are directly measured, while the magnet velocity~$\dot{s}$ has to be estimated.
The state estimation is achieved by utilizing the filtering approach proposed in~\cite{HanKim16}.
The underlying idea involves three simple linear filters which process the measured air gap signal~$s$ and the measured magnet acceleration~$\ddot{z}$ to estimate the magnet velocity~$\dot{z}$ and air gap velocity~$\dot{s}$ robustly.
This approach provides a good trade-off between noise suppression and estimation delay, and is well-suited for real-time implementation in embedded systems.
With this setup, the MPC design is completed and the focus shifts to the actual implementation and solution of the optimal control problem in the next section.

\section{Model Predictive Control Implementation}
\label{sec:mpc_implementation}
Concerning the implementation of optimal control problem~\eqref{eq:ocp_levitation}, there exist multiple approaches in literature.
One important distinction is between \textit{direct} and \textit{indirect} methods.
For a general overview the reader is referred to~\cite{RawlingsMayneDiehl17} or~\cite{DiehlEtAl06}.
In the following, the two methods are shortly introduced together with the used software toolkits.

\subsection{Direct Methods}
\label{sec:mpc_impl_direct}
Direct methods transform the continuous-time optimal control problem~\eqref{eq:ocp_levitation} into a finite-dimensional discrete-time nonlinear program, i.e., a nonlinear optimization problem in discrete variables.
There exist different discretization strategies for this purpose, the most common one being the so-called direct multiple shooting discretization.
The method originates from the field of boundary value problems, but can be easily adapted to optimal control problems~\cite{BockPlitt84}.
Thereby, the prediction horizon~$T$ is divided into $N$ shooting intervals whose boundaries are the shooting nodes.
The resulting discretization ratio is denoted with~$\delta_N = \nicefrac{T}{N}$.
Within each shooting interval the nonlinear system dynamics~$\dot{\bm{x}} = \bm{f}(\bm{x}, u)$ is discretized with a suitable numerical integration scheme, if no exact discretization can be given.
Hence, the integration over the whole prediction horizon is split up, which provides robustness, especially for unstable and highly nonlinear systems.
Furthermore, since input and state trajectories are simultaneously optimized, the sensitivity of the optimization parameters is much smaller, compared to, e.g., single shooting~\cite{DiehlEtAl06}.
To ensure that the resulting trajectories fulfill the original ODE, continuity constraints in between shooting intervals are defined.
The discretized optimal control problem results in a nonlinear program, which, however, shows a useful block-sparse structure.
Such a structure can be effectively exploited by state of the art optimal control problem solvers.

\begin{figure}[t]
	\centering
	\includegraphics{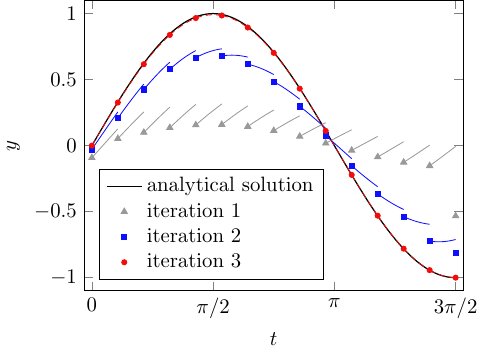}
	\caption{Illustration of the multiple shooting discretization for a simple boundary value problem.}
	\label{fig:04_multiple_shooting_demo}
\end{figure}

Figure~\ref{fig:04_multiple_shooting_demo} illustrates the multiple shooting method for a second-order linear system, where the analytical solution is the sine function.
Neither does the system dynamics contain inputs nor does an optimization take place.
Instead, simply the nonlinear equations given by the continuity constraints and boundary constraints are solved.
The solution trajectories for different iterations of the nonlinear equation solver are shown.
Once converged, the original ODE is fulfilled up to numerical precision and the analytical solution is recovered.

For an implementation of the arising optimal control problems in Matlab / Simulink the CasADi toolbox~\cite{AnderssonEtAl19} in combination with the nonlinear programming solver IPOPT~\cite{WaechterBiegler06} is used.
CasADi provides a symbolic manipulator with automatic differentiation capabilities, which is used to formulate the NLP, implementing the above-described discretization techniques in Matlab.
The interior point method IPOPT is then used in the background to compute a numerical solution of the NLP in each time step of the MPC algorithm.
Due to the complexity of the IPOPT solver, this approach is however not feasible for embedded application.
Furthermore, it is not tailored for fast computation or even real-time.

For embedded application, the \textit{acados} toolkit~\cite{VerschuerenEtAl22} is used instead, as a representative for a direct method.
The toolkit implements the direct multiple shooting method in combination with sequential quadratic programming (SQP) \cite{NocedalWright06} to solve the resulting nonlinear program.
Furthermore, it supports the real-time iteration (RTI) scheme proposed in~\cite{DiehlEtAl06}, which reduces the iterative SQP approach to the solution of one quadratic program (QP) per OCP solution.
The backend of acados provides interfaces to many popular QP solvers, e.g., HPIPM~\cite{FrisonEtAl20} or qpOASES~\cite{FerreauEtAl14}.

\subsection{Indirect methods}
\label{sec:mpc_impl_indirect}
Indirect methods take a different approach to solving the optimal control problem~\eqref{eq:ocp_levitation}.
Instead of discretizing the OCP first and subsequently solving the resulting finite-dimensional NLP, indirect methods tackle the OCP by first deriving the necessary conditions for optimality, typically using Pontryagin’s Maximum Principle (PMP), which transform the OCP into a continuous-time boundary value problem involving state, costate, and optimality conditions.
Other classical indirect approaches, such as those based on the Euler–Lagrange or Hamilton–Jacobi–Bellman formulations, follow a similar principle of deriving analytical optimality conditions before numerical solution, see~\cite{Kirk70} for a broader introduction.
\begin{table*}[b]
	\centering
	\caption{Comparison of direct and indirect methods for optimal control problem solution.}
	\renewcommand{\arraystretch}{1.2}
	\begin{tabularx}{\textwidth}{r|X|X}
		& \textbf{direct methods} & \textbf{indirect methods} \\
		\hline
		approach & discretize and solve NLP & derive continous time optimality conditions and solve numerically  \\
		constraint handling & straightforward via NLP formulation  & via projection and augmented Lagrangian method \\
		numerical robustness & numerically stable & numerically sensitive and potentially unstable\\
		real-time implementation & acados~\cite{VerschuerenEtAl22} & GRAMPC~\cite{EnglertEtAl19}
	\end{tabularx}
	\label{tab:direct_vs_indirect}
\end{table*}

Indirect methods offer the advantage of providing deep analytical insight into the structure of the optimal control problem by explicitly deriving necessary conditions for optimality. 
They can yield highly accurate solutions when the problem formulation and boundary conditions are well understood. 
However, these methods suffer from practical disadvantages, including sensitivity to initial guesses, numerical instability, and difficulties in handling state and input constraints. 
Consequently, indirect methods are less suited for large-scale or real-time control applications compared to direct methods~\cite{DiehlEtAl06}.

One popular approach to solving the resulting optimality conditions is the gradient projection method (GPM)~\cite{Rosen61}.
In this iterative scheme, the control input trajectory is updated in the direction of the negative gradient of the Hamiltonian with respect to the control input.
This corresponds to a gradient descent step, where the step size can be fixed or determined using a line search method.
After each update, a projection onto the admissible input constraint set is performed to maintain feasibility.
By performing repeated forward and backward integrations of the state and costate dynamics and successively refining the control input trajectory, the GPM iteratively converges toward a solution that satisfies both the optimality and feasibility conditions, stopping once a convergence criterion is met.

For the practical implementation of the gradient projection method, the GRAMPC toolbox~\cite{EnglertEtAl19} is used in this work. 
GRAMPC provides a tailored implementation of the GPM for embedded applications, including efficient numerical integration schemes and support for input and state constraints.
It is designed to be lightweight and suitable for real-time applications on embedded hardware.
It is worth noting that for the levitation control problem~\eqref{eq:ocp_levitation} only simple box constraints on the control input are imposed. 
Consequently, the application of more sophisticated constraint-handling techniques, such as the augmented Lagrangian framework proposed in~\cite{EnglertEtAl19}, is not required.
These equations are then exported as C code and integrated into the GRAMPC template structure. 
The GRAMPC algorithm can be tuned via several parameters, such as the maximum number of gradient decent steps and the integration step size~$\delta_N = \nicefrac{T}{N}$, where $T$ is the prediction horizon length and $N$ the number of integration steps.
Together with the convergence tolerances, the parameters are chosen to balance computational efficiency and solution accuracy according to the requirements.
If the computational resources are limited, the number of gradient descent steps can be reduced, which leads to faster computation times at the cost of solution accuracy, yielding a suboptimal control input.
This concludes the comparison of direct and indirect methods for solving optimal control problems, with a summary of the findings in Table~\ref{tab:direct_vs_indirect}.

\subsection{Embedded Implementation}
\label{sec:mpc_impl_embedded}
One aspect of this work is to investigate the applicability of the introduced model predictive control algorithms on embedded hardware, which is necessary for an implementation in the real system.
For research purposes, the AMD Zynq 7000 System-on-Chip (SoC), is chosen.
Although the chip features two $32\,\mrm{bit}$ ARM Cortex-A9 microprocessor cores as well as a field programmable gate array (FPGA), only one microprocessor core running with a clock frequency of $650\,\mrm{MHz}$ is used for the presented results.
Note that this microprocessor choice is not related in any sense to the real Maglev vehicle's control hardware and is purely chosen based on availability and convenience.

\begin{figure*}[t]
	\centering
	\includegraphics{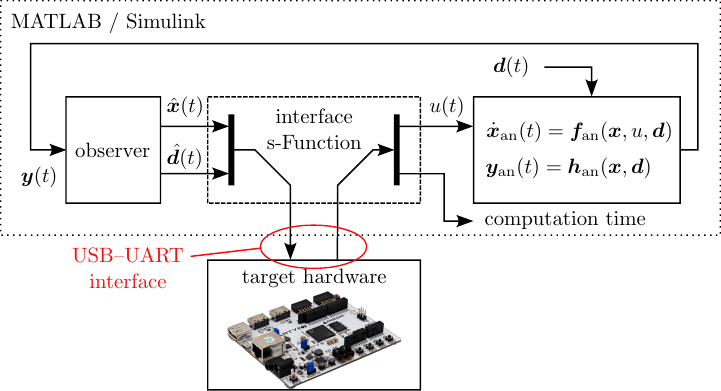}
	\caption{Processor-in-the-loop (PiL) setup including the simulation model implemented in Matlab / Simulink and the control algorithm implemented on the embedded target hardware.
	The communication is realized by a USB--UART interface.}
	\label{fig:pil_setup}
\end{figure*}

The control algorithms are exercised in a so-called processor-in-the-loop (PiL) environment, as it is sketched in Figure~\ref{fig:pil_setup}.
The system dynamics of the analysis model, i.e., Equations~\eqref{eq:analysis_model_dynamics} and \eqref{eq:analysis_model_output}, as well as the observer for state-feedback MPC are implemented in Matlab / Simulink and provide a simulation environment to test the controller.
The control algorithm itself is implemented on the microcontroller.
During simulation, communication between microcontroller and simulation model is realized by a USB--UART interface.
After every sampling interval, the estimated state and disturbance are sent to the microcontroller, which computes the next control input and measures the elapsed computation time.
Both values are sent back to the simulation model and the simulation continues for one sampling interval.

The implementation of the control algorithm on the microprocessor needs a short explanation.
Both toolboxes GRAMPC and acados provide convenient interfaces to Matlab, which enables fast prototyping and tests in simulation.
Using acados, the nonlinear system dynamics, cost function and constraints are encapsulated in symbolic expressions using the CasADi toolbox.
In the background CasADi's C code generation ability is then used to automatically generate a complete description of the controller in C.
The resulting C files can either be used to build a Simulink s-function, which is still part of the acados tool chain and used for simulation, or they can be manually compiled and linked into the control application for the embedded hardware.
The GRAMPC toolbox on the other hand does not provide automatic code generation, which leaves two possibilities.
One option is to take the same symbolically formulated expressions in CasADi, add needed gradients using its automatic differentiation functions and manually trigger the C export to incorporate the resulting code into the template provided by GRAMPC.
While this leads to a partially automated workflow, an analysis of the code showed a rather inefficient implementation, especially for embedded systems.
CasADi splits the computation into many small parts which leads to frequent calls to sub functions.
Furthermore, a specialty of the magnet model are its nonlinear coefficients stored in lookup tables, which again is only inefficiently implemented using CasADi's C export.
Hence, the needed ingredients for the GRAMPC toolbox are manually implemented in C, which results in a speed-up of approximately a factor of three.
For the acados toolbox, a manually optimized implementation of the system dynamics and cost function is not available at the time of writing this article.
However, since real-time capability of the discussed algorithms is not the main focus here, this is not critical for the presented results.

Finally, to implement the control algorithms, both toolboxes, GRAMPC and acados, themselves are compiled to static libraries specific to the ARM Cortex-A9 microprocessor.
This is done on a Linux machine with the cross-compiler provided by ARM and requires only little adaptation of the build workflow of the individual toolboxes.
All builds are compiled without debug symbols and with optimization level \texttt{-O2}.

\section{Simulative Investigation}
\label{sec:results}
This section provides a collection of insights gained either by pure simulation of the above-introduced closed loop or within the PiL setup described in Section~\ref{sec:mpc_impl_embedded}.
A few selected aspects are investigated in detail, e.g., how to choose the length of the prediction horizon~$T$ of the model predictive controllers and how the weighting factors of the MPC's cost function influence the closed-loop behavior.
Furthermore, some studies regarding the robustness of the MPC algorithms with respect to selected disturbances are carried out.
Finally, computation times and the quality of the obtained solutions are recorded for the two fast MPC schemes implemented on the AMD Zynq 7000 hardware platform.
Fast MPC usually compute suboptimal solutions to the optimal control problem.
Hence, this suboptimality is quantified with respect to fully converged OCP solutions.

\subsection{Prediction Horizon}
\label{sec:results_prediction_horizon}
The choice of the prediction horizon length~$T$ directly influences the closed-loop stability and performance of the MPC controller.
\begin{figure}[t]
	\centering
	\includegraphics{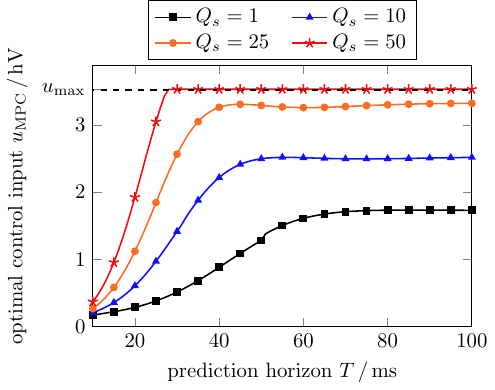}
	\caption{Comparison of the optimal input for different prediction horizon lengths and air gap weightings for the initial state $\bm{x}_0=[0.5, \; 0, \; 0.5]^\top$.}
	\label{fig:05_horizon_length_comparison}
\end{figure}
Since the OCP~\eqref{eq:ocp_levitation} does not include terminal costs or terminal constraints to formally guarantee closed-loop stability, the horizon length has to be chosen sufficiently long~\cite{RebleAllgoewer12,ScokaertMayneRawlings02}.
While such terminal ingredients guarantee stability for the nominal system, they do not hold in the presence of disturbances and model-plant mismatches.
Robust MPC approaches~\cite{BemporadMorari99} can be used to tackle these issues, but are not considered in this work.
Here, the goal is to find a minimum possible prediction horizon length by simulative investigations.
Figure~\ref{fig:05_horizon_length_comparison} shows the optimal control input~$u_\mrm{MPC}$ when solving the OCP~\eqref{eq:ocp_levitation} for different prediction horizon lengths and air gap weights for a given initial condition.
All other weighting parameters are held constant.
Notably, the optimal control input always converges for increasing prediction horizon lengths.
Furthermore, a large penalty on the air gap leads to more aggressive controller behavior, which is expected.

\noindent
However, once the optimal control input saturates at the constraint, increasing the horizon length has no further effect.
Thus, there is a trade-off between the prediction horizon length and the chosen cost function weights.
In the investigations to follow, a horizon length of $T=50\,\mathrm{ms}$ is used, as this value ensures convergence of the optimal control input for the considered air gap weights while avoiding unnecessary computational effort caused by longer horizons.
Note that this open-loop study only provides a first indication of a suitable prediction horizon length.
The closed-loop stability still has to be verified in simulative closed-loop investigations including disturbances, which is done in Section~\ref{sec:results_robustness}.

\subsection{Cost Function Parameterization}
\label{sec:results_cost_tuning}
As introduced in Section~\ref{sec:mpc_design}, each time the model predictive controller is invoked, the optimal control problem~\eqref{eq:ocp_levitation} is solved.
The question remains how to choose the weighting factors~$\bm{Q}$ and $R$ in the cost function~\eqref{eq:ocp_levitation_a}, apart from the theoretical requirements $\bm{Q} \succeq 0$ and $R > 0$.
Note that the complete objective function may be scaled by an arbitrary positive factor without changing the minimizer.
Hence, without loss of generality, set $R = 1$.
Further, note that the solution of OCP~\eqref{eq:ocp_levitation} is carried out with scaled variables, such that all outputs and the input take on values of similar magnitude.

\begin{figure}
	\centering
	\includegraphics{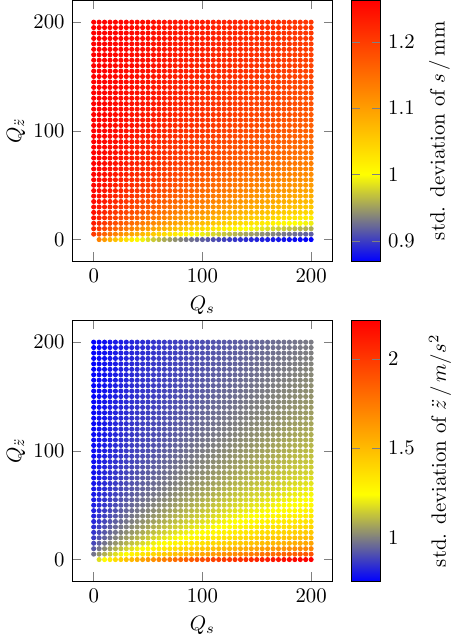}
	\caption{Standard deviations of air gap (left) and absolute magnet acceleration (right) depending on the cost function weighting factors~$Q_s$ and $Q_{\ddot{z}}$ with $Q_I = 5$ fixed.
	}
	\label{fig:results_cost_tuning}
\end{figure}

With these preliminaries, a small parameter study is performed to investigate the influence of the individual weighting factors in $\bm{Q} = \mrm{diag}(Q_s,\,Q_{\ddot{z}},\,Q_I)$.
The three factors are sampled on a regular grid of adequate ranges and closed-loop simulations of the half magnet model are performed under the influence of disturbances.
For evaluation, the standard deviation of the air gap signal and the absolute magnet acceleration signal are considered.
Small deviations from the reference air gap indicate good tracking, while a small magnet acceleration means that less vibrations are introduced into the system.
The studies reveal that the factor~$Q_I$ only has marginal influence on the closed-loop behavior.
Hence, Figure~\ref{fig:results_cost_tuning} shows the mentioned standard deviations for constant~$Q_I = 5$.
As expected, increasing the value of $Q_s$ leads to improved air gap tracking with the trade-off of higher magnet accelerations.
Increasing the penalty on the magnet acceleration has the opposite effect.
The only unstable combination of weighting factors in the investigated set is $Q_s = Q_{\ddot{z}} = 0$.
Based on this parameter study and if not mentioned otherwise, further simulations use the parameters $\bm{Q} = \mrm{diag}(75,\,15,\,5)$ and~$R = 1$, which is a reasonable compromise between air gap tracking and induced vibrations.
While this choice is suitable for simulations with the Transrapid vehicle parameters, an application to the real vehicle would require tuning based on experiments and lead to different cost function parameters.

\subsection{Robustness against Disturbances}
\label{sec:results_robustness}
Primarily, the magnetic levitation system is exposed to two disturbance sources, the bending motion and irregularity of the guideway~$d_{\mrm{gw}}$ which are discussed next, and the varying load~$F_{\mrm{load}}$ from the vehicle, caused by varying passenger number, but also aerodynamic effects.
In~\cite{SchmidEberhard21}, offset-free MPC schemes for the Maglev system are presented, which use disturbance estimation to mitigate these varying load disturbances.
Here, the focus is set on the guideway irregularities and how the designed MPC controller is able to handle them.

\begin{figure}[b]
	\centering
	\includegraphics{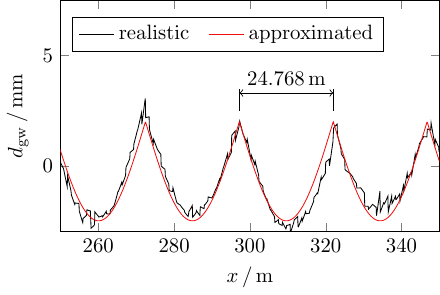}
	\caption{Approximated and realistic guideway models used in simulation.
	The girder length is fixed to $\lambda_{\mrm{gw}} = 24.768\,\mrm{m}$, the typical length is taken as suggested in \cite{EBA07en}.
	}
	\label{fig:results_gw_profile}
\end{figure}

The majority of the guideway is constructed using girders of equal length, which are placed on pillars.
Due to bending of the girders under the load of the vehicle, the control system is faced with a periodic disturbance which depends on the vehicle's speed~$v$. The disturbance frequency~$f_{\mrm{gw}}$ is given by
\begin{equation}
	f_{\mrm{gw}} = \frac{v}{\lambda_{\mrm{gw}}}
\end{equation}
with $\lambda_{\mrm{gw}}$ being the length of the guideway girders.
The guideway profile can be efficiently approximated by the function $d_{\mrm{gw}}(t) = \big| \mrm{sin}(\pi f_{\mrm{gw}} t) \big|$ as it is displayed in Figure~\ref{fig:results_gw_profile}.
Note that due to taking the absolute value the frequency has to be halved, such that the usual factor of two cancels out.
The figure also showcases a more detailed model of the guideway which further includes random offsets at the pillars, due to construction tolerances and unevenness caused by the stator packs' mounting.
A more detailed description on this guideway model is given in~\cite{ZhengEtAl18}.

\begin{figure}
	\centering
	\includegraphics{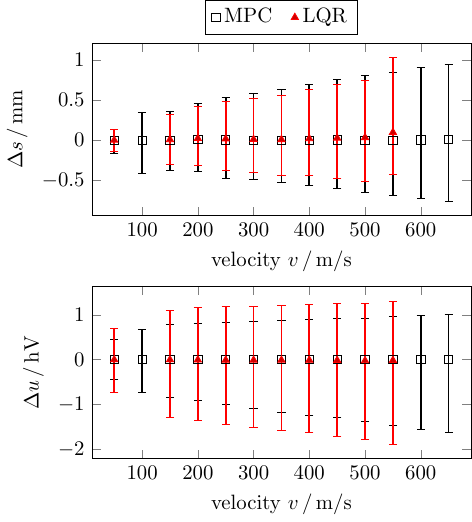}
	\caption{Closed-loop air gap and control input deviations from their nominal values for different velocities. The markers denote the mean value, while the error bars denote the standard deviation, respectively.}
	\label{fig:04_robustness_velocity_comparison}
\end{figure}

To investigate the robustness of the MPC controller against these guideway disturbances, closed-loop simulations with the half magnet analysis model~\eqref{eq:analysis_model_dynamics} are carried out for different velocities~$v$ ranging from $50\,\mrm{km/h}$ to $650\,\mrm{km/h}$ in steps of $50\,\mrm{km/h}$.
An LQR controller is utilized as a baseline for comparison.
In these simulations, the realistic guideway excitation is used and the closed-loop system is simulated for $20\,\mrm{seconds}$ at each velocity.
Figure~\ref{fig:04_robustness_velocity_comparison} shows the resulting mean control error as well as the standard deviation of the air gap and control input.
There, the markers denote the mean value over the simulation time, while the error bars denote the standard deviation of the respective signals.
Note that the standard deviations are computed separately for all values above and below the nominal values, in order to emphasize the asymmetry of the signals.
For example, the air gap's standard deviation is higher for values above zero for both controllers, which illustrates the fact that the air gap is more often larger than the desired value.
Notably, the LQR is not able to stabilize the system for velocities beyond $500\,\mrm{km/h}$ with the chosen parameterization, which is why these simulations are missing.
This is mainly caused by the input constraint, which the LQR is not able to handle and results in instability of the system.

Even at lower speeds around $100\,\mrm{km/h}$, the LQR can become unstable due to this limitation.
However, this does not necessarily imply that LQR as a method is uncapable for these speeds, but a reparameterization is needed, while this is not the case for MPC.
On the other hand, the MPC controller is able to stabilize the system for all considered velocities, thereby displaying its robustness against the guideway disturbance over a wide range of velocities and excitation frequencies.
This is even more impressive considering the very fast dynamics of the system at high velocities and the fact that the controller does not explicitly know about the disturbance, i.e., is not able to predict it.
However, this comes at the price of slightly higher standard deviations in the air gap control error.
The MPC is less aggressive compared to the LQR controller with the benefit of stability even at high velocities.
This is confirmed by reviewing the control input in Figure~\ref{fig:04_robustness_velocity_comparison}, where the LQR shows significantly higher standard deviations, thus confirming its more aggressive behavior.
The standard deviation of the MPC's control input also increases with higher velocities, since the controller has to react faster to the disturbance, but always remains below the LQR's values.
Concerning aspects like energy efficiency and durability of the system, MPC is advantageous in this case, since unnecessarily large electric currents are avoided.
This emphasizes the MPC controller's ability to perform well over a wide operation range, without the need for further adaptations compared to classical linear controllers such as an LQR.
Further improvements are expected once the guideway disturbance is explicitly considered in the MPC design, e.g., by using a disturbance model, preview information, or robust MPC techniques, which could be part of future work.

\subsection{Computation Times and Suboptimality}
\label{sec:results_pil}
Following the discussions about the optimal length of the prediction horizon, some further studies are carried out using the processor-in-the-loop (PiL) setup.
The focus is set to quantify the quality of the solutions to the MPC problem when using fast MPC algorithms.
Those algorithms only compute suboptimal solutions to the underlying OCP, to reduce the computational effort.
Ideally, the resulting controller behaves just as the original MPC with fully converged OCP solutions.
To investigate this aspect, simulations are carried out with the half magnet analysis model under the influence of external disturbances, namely the guideway irregularities discussed in Section~\ref{sec:results_robustness} at a speed of $430\,\mrm{km/h}$.
Quantification of the suboptimality is performed by two measures, namely the $L2$-norm of the control input signal as well as the relative cumulative suboptimality as used in~\cite{VerschuerenEtAl22}.
Let us define the closed-loop control input signal $u[k] := u_\mrm{MPC}(t_k)= u_\mrm{MPC}(k \delta)$ which results from applying the model predictive controller to the half magnet analysis model, i.e., closing the loop.
The signal is recorded from the continuous-time simulation with a fixed recording interval of~$\delta = 1\,\mrm{ms}$.
Then, the $L2$-norm of the control input signal of length~$K$ is defined as
\begin{equation}
	\Vert u \Vert :=
	\sqrt{\sum_{k = 1}^{K} u^2[k]}
	\;.
\end{equation}
It can be interpreted as the energy applied to the system accumulated over the simulation time~$K \delta$ and hence be used to compare different control algorithms for energy efficiency.

Another measure often used to evaluate an optimization based controller to a reference controller is the relative cumulative suboptimality (RCSO)~\cite{VerschuerenEtAl22}.
Let $u_{\mrm{ref}}$ and $u_{\mrm{test}}$ be the closed-loop control inputs recorded using the reference controller and a test controller, respectively.
Furthermore, the corresponding state signals~$\bm{x}_{\mrm{ref}}$ and~$\bm{x}_{\mrm{test}}$ are needed as well as an arbitrary cost function~$l(\bm{x}, u)$.
For convenience reasons the same cost function~\eqref{eq:ocp_levitation_a} as used in the MPC design is used also for evaluation, here.
Then, the RCSO accumulated up to simulation time $K \delta$ is defined as
\begin{equation}
	\text{RCSO}[K] :=
	\left|
	\frac{
		\displaystyle
		\sum_{k = 1}^{K} l(\bm{x}_{\mrm{test}}[k], u_{\mrm{test}}[k])
		-
		l(\bm{x}_{\mrm{ref}}[k], u_{\mrm{ref}}[k])
	}{
		\displaystyle
		\sum_{k = 1}^{K} l(\bm{x}_{\mrm{ref}}[k], u_{\mrm{ref}}[k])
	}
	\right|
	\;.
\end{equation}
The RCSO expresses how close the tested controller gets the reference controller by means of comparing the relative error measured by the cost function.
In the unrealistic ideal case, the RCSO value would be zero.
Small values indicate that there are only small differences in the closed-loop behavior.

\begin{figure}[h]
	\centering
	\includegraphics{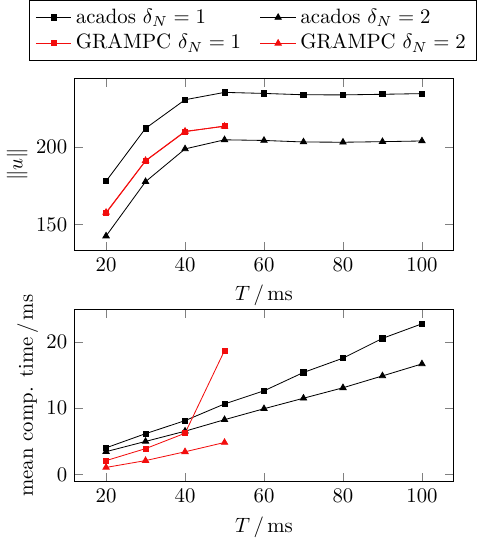}
	\caption{$L2$-norm of the closed-loop control input signal (left) and computation times measured on the embedded hardware (right) for different configurations of the model predictive controller.}
	\label{fig:results_energy_computation_time}
\end{figure}

Figure~\ref{fig:results_energy_computation_time} shows the obtained $L2$-norm of the control input signal for various controller configurations with respect to prediction horizons of different lengths~$T$.
Furthermore, the mean computation times for the acados and GRAMPC implementations are displayed as measured on the microcontroller.
For the different discretizations~$\delta_N$ of the prediction horizon, the reader is referred to their definition in Section~\ref{sec:mpc_impl_direct}, and Section~\ref{sec:mpc_impl_indirect}, respectively.
The MPC based on CasADi and IPOPT, which will be used as the fully converged reference controller in the following, is not implemented on the microcontroller, hence no computation times are recorded.
Concerning energy efficiency, the observations from Section~\ref{sec:results_prediction_horizon} reflect themselves again.
Up to prediction horizon lengths of approximately~$50\,\mrm{ms}$, the controller shows an increasingly aggressive 
behavior, saturating the control input, which increases the energy consumption.
Past $50\,\mrm{ms}$, only marginal changes in the consumed energy are visible.
Using the GRAMPC algorithm, the discretization of the prediction horizon, as long as it is kept in reasonable limits, does not influence outcome, which can be exploited to save computation time.
Using acados, the discretization, i.e., the number of shooting intervals seems to have a non-negligible influence on the control input where a coarser discretization leads to less aggressive behavior of the controller.
The observed computation times reflect what is already expected.
A longer prediction horizon or a finer discretization of the same prediction horizon, respectively, increases the number of floating point operations and therefore the computation time.

One detail about the GRAMPC algorithm has to be especially mentioned, namely stable simulations could only be achieved for~$T \leq 50\,\mrm{ms}$.
For longer prediction horizons the numerical solution to the optimal control problem~\eqref{eq:ocp_levitation} does not converge.
This is a general problem with indirect MPC methods, when they are implemented for highly nonlinear and unstable systems~\cite{DiehlEtAl06}.
Using the finely discretized prediction horizon, already for~$T = 50\,\mrm{ms}$, the convergence check has to be turned off, which results in the visible peak of the computation time.
The acados implementation, a representative of a direct MPC method, does not come with this problem, such that the prediction horizon could be extended far beyond $100\,\mrm{ms}$.

\begin{figure}[t]
	\centering
	\includegraphics{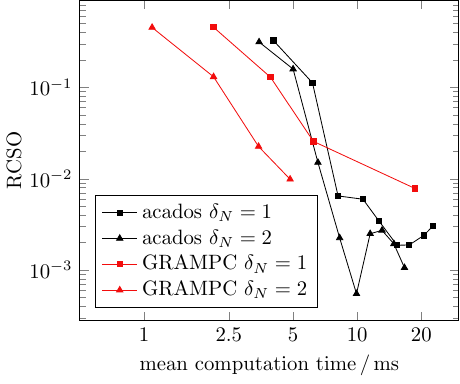}
	\caption{Relative cumulative suboptimality (RCSO) achieved by the acados and GRAMPC algorithms for different configurations with the fully converged MPC using CasADi and IPOPT as a reference.
	For each method different lengths for the prediction horizon are tested.
	The reference controller uses~$T = 100\,\mrm{ms}$ and $\delta_N = 1$.}
	\label{fig:results_rcso}
\end{figure}
Figure~\ref{fig:results_rcso} relates the obtained RCSO of the tested model predictive controllers to the observed mean computation time on the embedded hardware.
Simulations are performed with the same values for the prediction horizon as before, starting at $20\,\mrm{ms}$, and increasing the horizon in steps of~$10\,\mrm{ms}$.
The reference MPC implementation with IPOPT as underlying optimizer uses a prediction horizon with length~$T = 100\,\mrm{ms}$ discretized with~$\delta_N = 1$.

The general trend reveals that a longer prediction horizon leads to lower RCSO values, again with only small improvements for prediction horizons $T > 50\,\mrm{ms}$.
Comparing the two investigated discretization options, especially for the GRAMPC algorithm, one can effectively save computation time, with only marginal loss of accuracy as of the RCSO value.
For acados the same holds, but the effect is not as drastic.

Generally, the acados toolbox is not particularly optimized for embedded usage.
This does not only concern computation time, but memory consumption as well, with random access memory (RAM) usage of more than $1.5\,\mrm{MB}$, which is rather high for an embedded system.
All in all the results show that an embedded implementation is possible.
However, further adjustments to the algorithms or simplifications to the control model would have to be done to reach the real-time barrier of $\delta = 1\,\mrm{ms}$ for the Maglev system.

\section{Conclusion and Outlook}
\label{sec:conclusion}

Embedded nonlinear model predictive control is a promising approach for EMS-type Maglev systems, offering clear advantages over classical linear controllers. 
The results demonstrate that MPC outperforms linear approaches at high vehicle speeds and under guideway disturbances, enabling a wider operational range without further adaptations. 
Both direct and indirect fast MPC methods are shown to be feasible for real-time embedded implementation, but challenges for achieving computation times below $1\,\mrm{ms}$ remain.
Simulation and processor-in-the-loop (PiL) studies further confirm the practical viability of embedded MPC on representative test hardware.

Looking ahead, several opportunities for further enhancements emerge from this work. 
One major direction is the development of advanced control strategies that can explicitly address the persistent guideway disturbances inherent to the presented Maglev system. 
This could involve adaptive control techniques, leveraging preview information about the track, or robust MPC methods to enhance disturbance rejection and overall robustness.
Therefore, ensuring that the control law remains dependable even in the face of model uncertainties and external disturbances will be essential as these advanced strategies are developed.

Another important direction is the continued optimization of embedded computation times. 
Achieving even faster and more efficient real-time control may be possible through the use of explicit MPC formulations. 
Such improvements would help push the boundaries of what is feasible on resource-constrained embedded hardware.
Addressing these challenges will be key to unlocking the full potential of embedded MPC for future high-speed Maglev applications.

%
%




\begin{IEEEbiography}[{\includegraphics[width=1in,height=1.25in,clip,keepaspectratio]{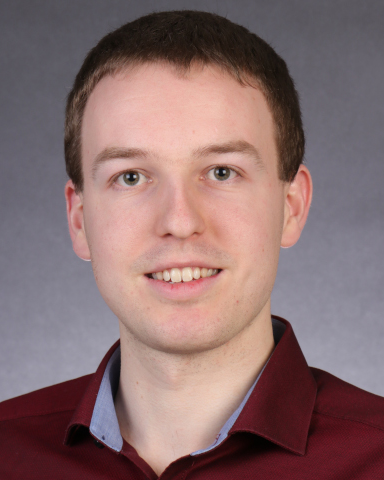}}]{Arnim Kargl}
received the BSc and MSc degrees in engineering cybernetics from the University of Stuttgart, Germany, in 2020 and 2022, respectively.
He is currently pursuing the doctoral degree with the Institute of Engineering and Computational Mechanics.
His research interests include modeling, simulation, and control of high-speed Maglev systems.
The focus hereby lies on embedded real-time control concepts.
\end{IEEEbiography}
\begin{IEEEbiography}[{\includegraphics[width=1in,height=1.25in,clip,keepaspectratio]{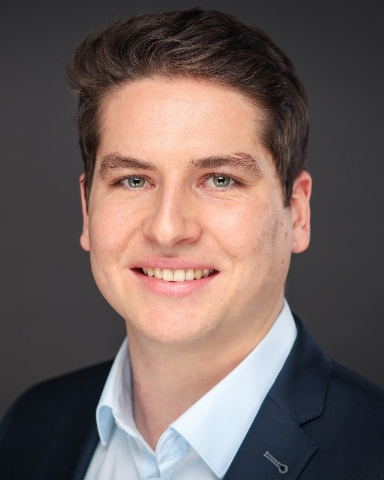}}]{Mario Hermle}
received the BSc and MSc degrees in engineering cybernetics from the University of Stuttgart, Germany, in 2019 and 2022, respectively.
He is currently pursuing the doctoral degree with the Institute of Engineering and Computational Mechan-
ics.
His research focuses on the development of Model Predictive Control (MPC) methods to enhance safety, stability, and ride comfort of high-speed Maglev vehicles.
\end{IEEEbiography}
\begin{IEEEbiography}[{\includegraphics[width=1in,height=1.25in,clip,keepaspectratio]{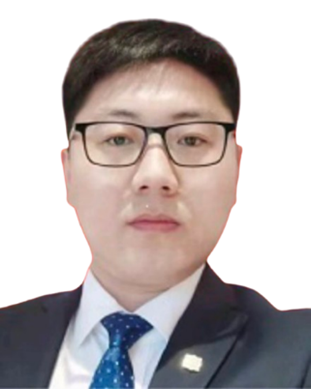}}]{Zhiqiang Zhang}
received his B.S., M.S., and Ph.D. degrees in Electrical Machines and Apparatus from Harbin University of Science and Technology in 2005, 2008, and 2013, respectively.
He is currently a professor-level senior engineer at the State Key Laboratory of High-speed Maglev Transportation Technology, CRRC Qingdao Sifang Co., Ltd., Qingdao, China.
His primary research interests include high-speed Maglev system integration, traction power supply, and levitation and guidance control technologies.
\end{IEEEbiography}
\begin{IEEEbiography}[{\includegraphics[width=1in,height=1.25in,clip,keepaspectratio]{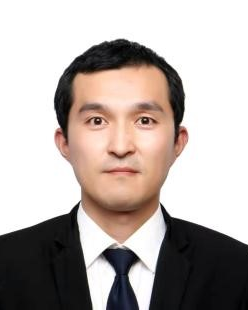}}]{Yanmin Li}
received the bachelor's degree in automation from Northeastern University in 2010, and the master's degree in Pattern Recognition and Intelligent Systems from the University of Chinese Academy of Sciences in 2017.
Currently, he is with the State Key Laboratory of High-speed Maglev Transportation Technology, CRRC Qingdao Sifang Co., Ltd., Qingdao, China.
His main research fields include levitation and guidance control as well as location technology for high-speed Maglev trains.
\end{IEEEbiography}
\begin{IEEEbiography}[{\includegraphics[width=1in,height=1.25in,clip,keepaspectratio]{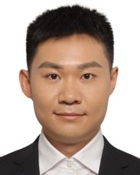}}]{Dainan Zhao}
received his B.Eng. degree in electrical engineering and electronics from the University of Liverpool in 2015, and his M.Sc. degree in electrical engineering and information technology from the University of Stuttgart in 2019.
He is currently an engineer at the State Key Laboratory of High-speed Maglev Transportation Technology, CRRC Qingdao Sifang Co., Ltd.
His research focuses on levitation and guidance control as well as power supply technologies for high-speed Maglev trains.
\end{IEEEbiography}
\begin{IEEEbiography}[{\includegraphics[width=1in,height=1.25in,clip,keepaspectratio]{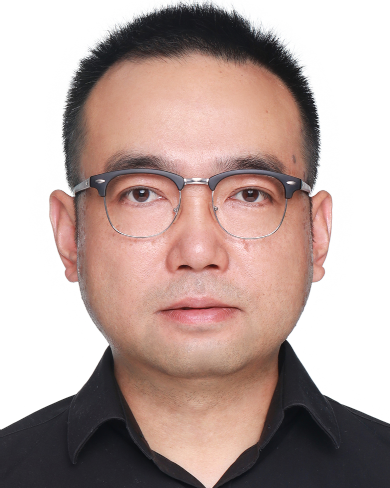}}]{Yong Cui}
is the executive director of the Chinese-German Research and Development Centre for Railway and Transportation Technology in Stuttgart and a professor at Hefei University.
Since 2020, he has also served as an adjunct professor at the University of Stuttgart.
His main professional fields are railway operation and control, simulation and optimization, as well as scheduling and planning.
\end{IEEEbiography}
\begin{IEEEbiography}[{\includegraphics[width=1in,height=1.25in,clip,keepaspectratio]{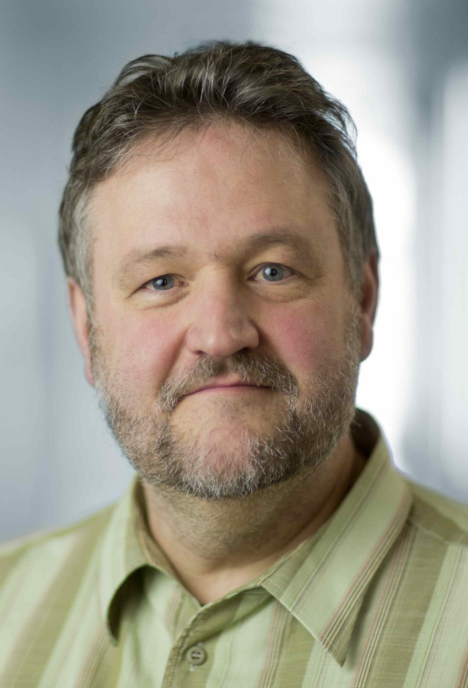}}]{Peter Eberhard}
is a full professor and since 2002 director of the Institute of Engineering and Computational Mechanics (ITM) at the University of Stuttgart, Germany.
He was Treasurer and Bureau member of IUTAM, the International Union of Theoretical and Applied Mechanics, and served before in many national and international organizations, for example, as Chairman of the IMSD (International Association for Multibody System Dynamics) or DEKOMECH (German Committee for Mechanics).
He is a member of the International Maglev Board.
With his team, he is interested in multibody dynamics, control, meshless methods, uncertainties, and other related areas.
\end{IEEEbiography}

\end{document}